\documentclass[showpacs,aps,prx,amsmath,superscriptaddress,twocolumn]{revtex4-1}

\usepackage{mathrsfs}
\usepackage{amssymb}
\usepackage{amstext}
\usepackage{txfonts}
\usepackage{mathtools}
\usepackage[pdftex]{color}
\usepackage[english]{babel}
\usepackage{graphicx}
\usepackage{bm}
\usepackage{float}
\usepackage{ulem}

\newcommand{\bra}[1]{\left\langle #1 \right|}
\newcommand{\ket}[1]{\left|#1\right\rangle}

\begin{document}

\title{Optimizing Variational Quantum Algorithms using Pontryagin's Minimum Principle}

\author{Zhi-Cheng Yang}

\affiliation{Physics Department, Boston University, Boston, Massachusetts 02215, USA}

\author{Armin Rahmani}

\affiliation{Department of Physics and Astronomy and Quantum Matter Institute, University of British Columbia, Vancouver, British Columbia, Canada V6T 1Z4}

\affiliation{Department of Physics and Astronomy, Western Washington University, 516 High Street, Bellingham, Washington 98225, USA}

\author{Alireza Shabani}

\affiliation{Google Inc., Venice, California 90291, USA}

\author{Hartmut Neven}

\affiliation{Google Inc., Venice, California 90291, USA}

\author{Claudio Chamon}

\affiliation{Physics Department, Boston University, Boston, Massachusetts 02215, USA}

\date{\today}

\begin{abstract}
We use Pontryagin's minimum principle to optimize variational quantum
algorithms. We show that for a fixed computation time, the optimal
evolution has a bang-bang (square pulse) form, both for closed and
open quantum systems with Markovian decoherence. Our findings support
the choice of evolution ansatz in the recently proposed Quantum
Approximate Optimization Algorithm. Focusing on the
Sherrington-Kirkpatrick spin-glass as an example, we find a
system-size independent distribution of the duration of pulses, with
characteristic time scale set by the inverse of the coupling constants
in the Hamiltonian. The optimality of the bang-bang protocols and the
characteristic time scale of the pulses provide an efficient
parameterization of the protocol and inform the search for effective
hybrid (classical and quantum) schemes for tackling combinatorial
optimization problems. For the particular systems we study, we find
numerically that the optimal nonadiabatic bang-bang protocols
outperform conventional quantum annealing in the presence of weak
white additive external noise and weak coupling to a thermal bath
modeled with the Redfield master equation.

\end{abstract}


\maketitle

\section{Introduction}
\label{sec:intro}

Quantum Annealing (QA) aims to solve computational problems by using a
guided quantum drive. The dynamics is generated by a time-dependent
Hamiltonian along a trajectory that ends at a final target Hamiltonian
whose ground state contains the solution of the
problem~\cite{Nishimori, Sipser, QAA}. QA is based on the adiabatic
theorem, which guarantees that if the Hamiltonian is changed
sufficiently slowly, transitions to excited states are suppressed
during the adiabatic evolution, thus preparing states that are close
to the target ground state. Unfortunately, the adiabatic condition
that ensures that the system remains in the instantaneous ground state
leads to long time scales for the solution of hard computational
problem. Within the framework of adiabatic computation, there has been
several theoretical proposals on the optimizations of the Quantum
Adiabatic Algorithm (QAA), such as heuristic guesses for the initial
state~\cite{Perdomo}, increasing the minimum gap~\cite{Mohan1,zhuang},
and the quantum adiabatic brachistochrone
formulation~\cite{Rezakhani}.

\begin{figure}[!ht]
\centering
\includegraphics[angle=0,origin=c,width=8cm]{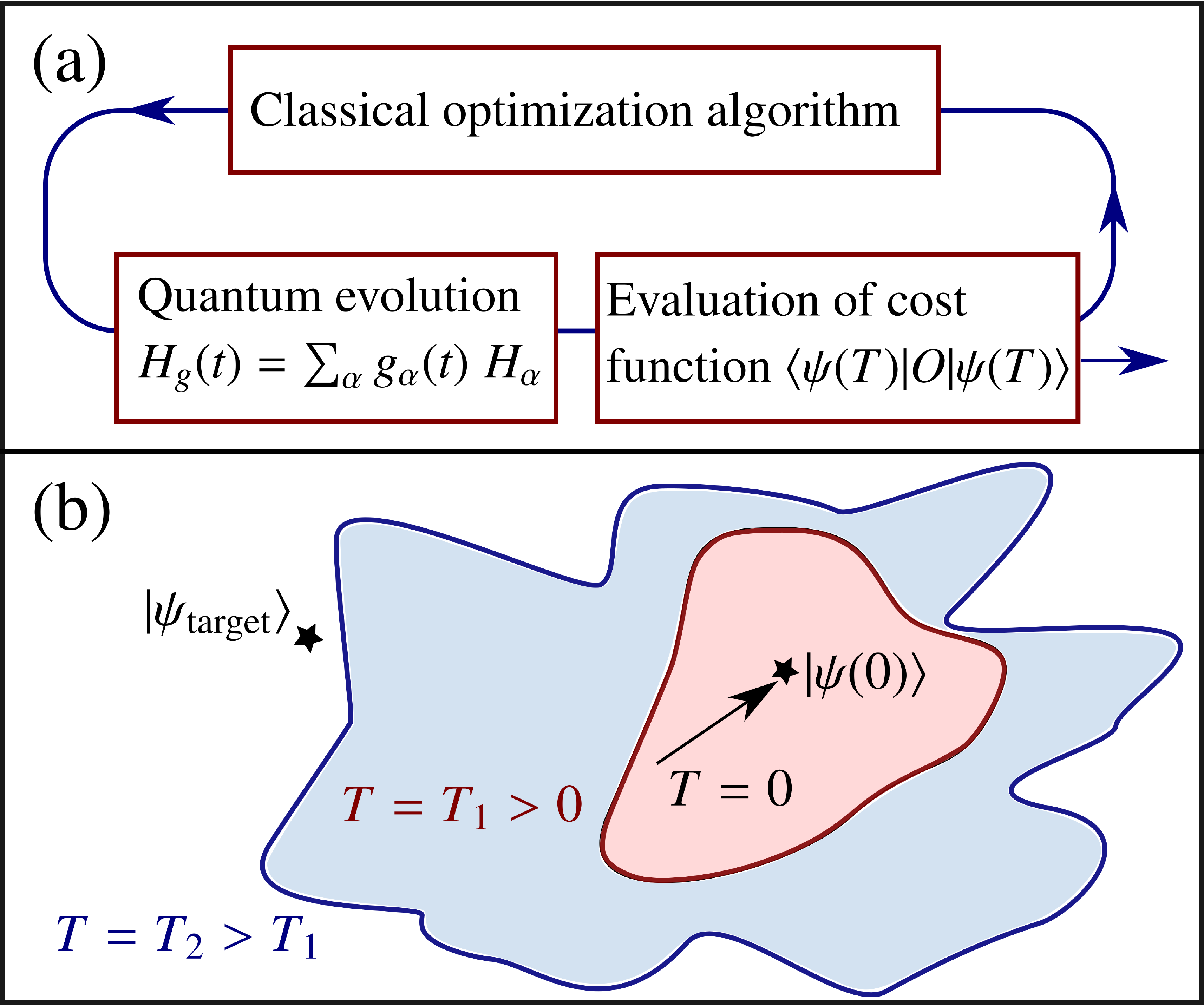}
\caption{(a) Variational quantum algorithm as a closed-loop learning
  control problem. (b) Increasing the total time expands the set of
  final states that one can reach with the variational protocols. The
  optimal protocol for a given time generates the closest state to a
  low energy target state within this reachable set.}
\label{fig:VQA}
\end{figure}

The adiabatic trajectory is not the only path for reaching the ground
state of a final Hamiltonian that encodes the solution of the
computational problem. More generally, one could imagine many other
paths, including those where the Hamiltonian is varied rapidly, that land at the desired state or, of practical interest, reach low energy states. In fact, it has been already found
that for certain hard instances of problems, fast nonadiabatic paths
can sometimes prevent the system from getting stuck at local minima, thus improve the search results~\cite{troyer, heim, crosson}. The
Variational Quantum Algorithm (VQA) is an example where one searches
for such possible paths, using optimization of the outcome via the
variation of a fixed number of parameters in the protocol. A hybrid machine, combining
classical optimization and quantum evolution, optimizes the
variational parameters. Such hybrid variational approaches have proved
useful in the context of quantum state
preparation~\cite{Rohringer,Rosi,Rahmani1,Rahmani2}. Recently,
Ref. \cite{Peruzzo:14} introduced a variational quantum eigensolver (VQE)
for applications in quantum chemistry. This idea was further explored
in \cite{Yung:14, Wecker:15, wecker, McClean:16, McClean:16-2} and
experimentally tested in \cite{Shen:15, Eichler:15, Omalley:15}. In a
related approach \cite{Farhi,Farhi:2,Farhi:3, Yechao}, Farhi \textit{et al.}
introduced a Quantum Approximate Optimization Algorithm (QAOA) for
combinatorial optimization problems based on a parameterized square-pulse ansatz
for dynamical evolution of the solver.

In this paper, we make a connection between VQA and optimal control
theory~\cite{control1, Pontryagin, control2, Mohan2}. VQA is essentially an adaptive feedback control \cite{adaptive, Brif:10}
of a quantum system with the objective function encoding the solution
of a computational problem, see Fig. \ref{fig:VQA}(a). It utilizes a
hybrid system comprised of a classical computer that searches for the
optimal variational protocol using measurements done on a quantum
machine that generates the final states corresponding to different variational protocols,
via a closed-loop learning method \cite{Dong:10}.

Using Pontryagin's minimum principle of optimal control, we show that
the optimal protocol for VQA has a ``bang-bang'' form. Our results put
the bang-bang ansatz of QAOA on a rigorous ground in contrast to VQA
with a continuous-time evolution.  A comparison of the performance of
the optimal nonadiabatic bang-bang protocol with conventional (linear
ramp) QAA demonstrates that the former significantly reduces error in
the final state in the absence of noise or decoherence. The advantage
over the linear ramp protocol in QAA survives weak dephasing white
noise as well as weak coupling to a thermal bath. Furthermore, we
perform a quantitative analysis of the characteristics of these
optimal protocols. We numerically find a system-size independent
distribution function for the duration of individual pulses, which may
facilitate the development of effective algorithms for the classical
optimizer through an efficient representation of the protocol with few
variational parameters. Interestingly, each of the pulses in our
bang-bang protocols contains commuting (either one-qubit or two-qubit)
terms. Thus our protocol can be implemented by applying a sequence of
one-qubit gates [ generated by the initial Hamiltonian, $g=0$ in
  Eq.~\eqref{eq1}] and two-qubit gates generated by the problem
Hamiltonian ($g=1$).

\section{Variational Quantum Algorithm}
\label{sec:VQA}

 Consider a computational optimization problem such as finding a sequence of $N$ bits that minimizes a certain function of all of the bits. To solve this problem with VQA, we consider a system of $N$ qubits with a parameterized Hamiltonian
\begin{equation}\label{eq:H_theta}
H_g(t)=\sum_\alpha g_\alpha(t) \; H_\alpha.
\end{equation}
Generically, we can cast the problem into generating a state $|\psi\rangle$ that minimizes a certain cost function such as the expectation value of an operator $O$ with respect to $|\psi\rangle$. A common example is finding the ground state of a disorderd classical Ising Hamiltonian \cite{Lucas:14}, where the operator $O$ is a Hamiltonian diagonal in the computational basis. In the context of quantum chemistry, VQE considers the operator $O$ to be the Hamiltonian of a molecule \cite{Yung:14}.

The essence of VQA, as depicted in Fig. \ref{fig:VQA}, is finding the time-dependent parameters $g_\alpha(t)$ over a time period $T$ such that 
 \begin{equation}
\ket{\psi(T)}={\cal T}{\rm e}^{-i\int_0^T H_g(t) dt}\ket{\psi(0)},
\label{eq:general}
\end{equation}
minimizes a cost function $\langle\psi(T)|O|\psi(T)\rangle$. Generically, the controls $g_\alpha(t)$ belong to a permissible set determined by the experimental setup. A common such set is given by simple bounds as seen in Eq.~\eqref{g-limits} below. The ideal solution could be a unique state $|\psi_{\rm target}\rangle$ [as depicted in Fig.\ref{fig:VQA}(b)] that is the ground state of the target Hamiltonian or more generally a set of states in the Hilbert space with an optimal figure of merit. One can either fix the initial state $|\psi(0)\rangle$ or add it to the list of the variational parameters (here we fix it motivated by experimental constraints). Generally, the longer the total time $T$, the closer we can get to an ideal solution.

One way to view this is to consider the reachable set, i.e., the set of all the final states one can reach by using one of the infinite number of permissible controls. The reachable set, naturally, grows with $T$ (in fact, if $g_\alpha=0$ is allowed, the reachable set for $T=T_1$ is strictly a subset of the reachable set for $T=T_2>T_1$).
As seen in Fig. \ref{fig:VQA}(b), there could be a critical time beyond which the reachable set includes the target state and an exact solution is possible. There is no advantage in increasing $T$ beyond this critical time. Generically, for smaller $T$, where the reachable set does not include the target state, the optimal protocols are highly constrained as they should prepare the closest point(s) of the reachable set to the target. For times longer than the critical time mentioned above, we expect an infinite number of protocols to produce the target as the evolution has extra time to meander in the Hilbert space. Our strategy is to fix $T$ and find the best variational protocol $g_\alpha(t)$. If the solution is not acceptable, we can increase $T$. Next we discuss how Pontryagin's minimum principle from optimal control theory determines the form of optimal $g_\alpha(t)$ functions.

\section{Pontryagin's minimum principle applied to VQA}
\label{sec:pontryagin}
\subsection{Bang-bang optimal protocols}\label{sec:bang}
 The parameters in Hamiltonian~\eqref{eq:H_theta} are generically constrained by their range:
\begin{equation}
g^{\min}_\alpha\leqslant g_\alpha(t)\leqslant g^{\max}_\alpha
\label{g-limits}
\end{equation}
during the evolution $0<t<T$. Eq. ~\eqref{g-limits} implies that, by assumption, each $g_\alpha$ can be tunned in the above range independently of the values of the other control parameters. For fixed initial state $|\psi(0)\rangle$, the coupling constants $g_\alpha(t)$ uniquely determine the final wave function. Consequently, the cost function, which we take as an arbitrary function of the final state, is a functional of $g_\alpha(t)$
\begin{equation}\label{eq:f_cost}
F[\left\{ g_\alpha(t)\right\}]={\cal F}(|\psi(T)\rangle).
\end{equation}

The Pontryagin's minimum principle~\cite{Pontryagin} is directly applicable here. Briefly, this theorem states that for a set of dynamical variables $\bm x$ evolving from given initial values ${\bm x}(0)$ with the equations of motions $\dot {\bm x}={\bm f}({\bm x}, \boldsymbol{  g})$, where $\boldsymbol{  g}$ are a set of control functions, the control functions $\boldsymbol{  g}^*$ that minimize an arbitrary function ${\cal F}[{\bm x}(T)]$ of the final values of the dynamical variable satisfy
\begin{equation}\label{eq:pont}
{\cal H}({\bm x}^*,{\bm p}^*,\boldsymbol{  g}^*)=\min_{\boldsymbol{  g}}{\cal H}({\bm x}^*,{\bm p}^*,\boldsymbol{  g})
\end{equation}
at any point in time and for each of the control functions. The optimal-control Hamiltonian is defined as ${\cal H}({\bm x},{\bm p},\boldsymbol{  g})\equiv {\bm f}({\bm x}, \boldsymbol{  g})\cdot{\bm p}$ for conjugate variables $\bm p$ that evolve as $\dot {\bm p}=-\partial_{{\bm x}}{\cal H}$ with boundary conditions ${\bm p}(T)=\partial_{{\bm x}}{\cal F}[{\bm x}(T)]$. Here the ``$*$'' superscript indicates the optimal solution corresponding to $\boldsymbol{  g}^*$.

An important consequence of Eq.~\eqref{eq:pont} is that if the equations of motion for $\bm x$, and consequently the optimal-control Hamiltonian $\cal H$, are linear in $ \boldsymbol{  g}$, generically, the optimal protocol is bang-bang, i.e., at any time during the evolution we have $g^*_\alpha(t)= g^{\min}_\alpha$ or $g^*_\alpha(t)= g^{\max}_\alpha$. This follows from the fact that at any point in time we need to choose $g_\alpha$ to minimize ${\cal H}({\bm x}^*,{\bm p}^*,\boldsymbol{  g})$. If the sign of the coefficient of $g_\alpha$ in the optimal-control Hamiltonian is positive (negative), we should then choose the smallest (largest) $g_\alpha$ from the permissible range~\eqref{g-limits}. In other words, the optimal protocol for each control function involves a sequence of sudden jumps between its minimum and maximum permissible values. The only caveat for the above argument is the possibility that the coefficient of a particular $g_\alpha$ in ${\cal H}({\bm x}^*,{\bm p}^*,\boldsymbol{  g})$ vanishes over a finite interval (since the sign of this coefficient determines whether we should choose the minimum or maximum value). We expect this special scenario to be nongeneric particularly for the disordered systems considered in the present paper.

In the quantum mechanical context, if the physical Hamiltonian is linear in the controls, the equations of motion and consequently the optimal-control Hamiltonian will also be linear, giving rise to bang-bang protocols as verified in several recent studies on optimal topological quantum computing~\cite{Rahmani3,Rahmani4}.
To find the protocol $\boldsymbol{  g}$ that minimizes the cost function in our case, we expand the wave function in a complete orthonormal basis, e.g., the computational basis $|z\rangle$ as $|\psi(t)\rangle=\sum_z A_z(t) |z\rangle $ and treat the real and imaginary parts of the amplitudes $A_z(t)$ as dynamical variables, which evolve according to the Schr\"odinger equation
\begin{eqnarray}
\dot{A}^R_z&=&{1\over 2}\sum_{\alpha,z'} g_\alpha
\left[\left(H_\alpha^{zz'}+H_\alpha^{z'z}\right)A^I_{z'}
-i\left(H_\alpha^{zz'}-H_\alpha^{z'z}\right)A^R_{z'}
\right],\\
\dot{A}^I_z&=&{1\over 2}\sum_{\alpha,z'} g_\alpha
\left[-\left(H_\alpha^{zz'}+H_\alpha^{z'z}\right)A^R_{z'}
-i\left(H_\alpha^{zz'}-H_\alpha^{z'z}\right)A^I_{z'}
\right],
\end{eqnarray}
where $H_\alpha^{zz'}\equiv \langle z|H_\alpha|z'\rangle$ and $A_z^{R,I}\equiv{\rm Re,Im}(A_z)$. Clearly, these equations of motion are linear in the control functions $g_\alpha$ and the cost function~\eqref{eq:f_cost} is a function of only the final values of the dynamical variables. Thus, the argument above holds and the optimal protocol is generically bang-bang regardless of the number of variational parameters. We remark that our optimal bang-bang protocol is nonadiabatic by construction, and we put no constraint on maximizing the degree of adiabaticity. The value of this result hinges upon the time scale over which a coupling constant is held fixed. The longer this time scale, the fewer parameters (switching times) 
are needed to represent the protocol. In fact, in the limit where this time scale goes to zero, any protocol can be approximated by a sequence of square pulses through Trotterization. In this paper, we find that the time scale above is indeed finite and is set by the energy scale of the Hamiltonian for the Sherrington-Kirkpatrick (SK) Ising spin-glass model [see Eq.~\eqref{eq1}].
\subsection{Presence of decoherence}

From a practical point of view, it is important to assess the validity of the closed system findings in the presence of decoherence. Again, a straightforward application of the Pontryagin principle extends the above results for a closed system evolution to an open quantum system with Markovian dynamics described by a Lindblad equation
\begin{equation}
\frac{d\rho}{dt}=-i\left[\sum_\alpha g_\alpha(t) \; H_\alpha, \ \rho\right]+\sum_\beta f_\beta(t)\left(2F_\beta \ \rho \  F^\dagger_\beta-\left\{F^\dagger_\beta F_\beta,\ \rho\right\}\right)
\label{eqLindblad}
\end{equation}
where the optimal protocol $\{g_\alpha(t),f_\beta(t)\}$, if controllable, are of type bang-bang. This is due to the linearity of the dynamical equation (\ref{eqLindblad}). A decoherence operator $F_\beta$ can represent either noise in the Hamiltonian parameters (in which case, $F_\beta$ is Hermitian~\cite{Pilcher2013,Rahmani2015}) or an engineered bath~\cite{open}. In the former case, $f_\beta$'s are constant rates of noise processes and in the latter, $f_\beta(t)$'s are control knobs that the Pontryagin's minimum principle says should vary in bang-bang form for an optimal protocol. In the rest of the paper, we only focus on closed systems Schr\"odinger dynamics when finding the optimal protocol. We do, however, discuss the effects of noise and open-system dynamics on our optimal protocols.

\section{VQA for the SK Spin-Glass Model}
\label{sec:model}

We now focus on a canonical problem in combinatorial optimization, namely the SK Ising spin glass \cite{SK} with the energy function 
\begin{equation}
C= \frac{1}{\sqrt{n}}\sum_{i,j=1}^n J_{ij}\sigma^z_i\sigma^z_j+\sum_{i=1}^n h_i \sigma^z_i.
\label{eq1}
\end{equation}
where $J_{ij}$ and $h_i$ are independent Gaussian random variables with zero mean and variance $J^2=h^2=1$, and each $\sigma^z$ spin can take the values $\pm 1$. The goal is to minimize $C$ over all the $2^n$ spin configurations. A multitude of practical combinatorial optimization problems map to this model. The computational cost of finding the minimum with classical algorithms is exponential in $n$.

In analogy with the simple instances of quantum annealing, we focus on
the case with only one control function $g(t)$ and use the following
parameterized Hamiltonian:
 \begin{equation}\label{eq2}
H_ g(t)=  g(t)C+[1- g(t)]B,
\end{equation}
with the operator $B\equiv -\sum_{i=1}^n \sigma^x_i$ representing a transverse field, which generates quantum fluctuations.

For the initial state, we choose the ground states of $B$. It is easy to prepare product state $|\psi(0)\rangle=\prod_i
\left(|\uparrow\rangle_i+|\downarrow\rangle_i\over \sqrt{2}\right)$ commonly used in other schemes such as the QAA. Here $\sigma^z_i|\uparrow\rangle_i=|\uparrow\rangle_i$ and $\sigma^z_i|\downarrow\rangle_i=-|\downarrow\rangle_i$. We would like to minimize the cost function $\langle\psi(T)|C|\psi(T)\rangle$. In the adiabatic scheme, a smooth ramp such as $g(t)={t\over T}$ is applied for $0<t<T$ and we can generate large overlap with the ground state of $C$ in the limit of large $T$. Here, we allow for arbitrary time dependence of the control function in the fixed range $0\leqslant g(t)\leqslant1$. According to the general argument of Sec.~\ref{sec:bang}, the optimal solution is bang-bang.

As discussed in the introduction, in VQA a classical optimization
algorithm commands a quantum system to find the optimal protocol
variationally from measurement of the cost function for many
protocols. This requires many repetitions and it is to our advantage
to use the shortest possible time $T$ for which the final state has an
acceptable overlap with the ground state of $C$ (projective
measurement is ultimately used in generating the ground state). In the
adiabatic scheme, we only need one shot but there are important
restrictions from the small energy gaps along the adiabatic
trajectory, which can lead to exceedingly long time scales, over which
quantum coherence cannot be even approximately sustained. Furthermore,
the presence of noise or modulation in the control fields places
important limitations on adiabatic schemes due to the emergence of the
recently proposed noise-induced anti-adiabaticity in the long-time
limit~\cite{Dutta}. Unlike QAA, which relies on the adiabatic theorem,
VQA has no known connection to instantaneous ground states and the
minimum gap to excitations as transitions to excited states during the
time evolution are allowed as long as the system eventually lands at
the ground state of the final Hamiltonian.

Given  the limitations of adiabatic scheme, a quantum approximate algorithm has been introduced for solving combinatorial optimization problems~\cite{Farhi, Farhi:2, Farhi:3, Yechao} in the spirit of VQA. The algorithm of Ref.~\cite{Farhi} uses an ansatz 
\begin{equation}
\ket{\boldsymbol{\gamma}, \boldsymbol{\beta}} = U (B, \beta_p) \, U (C, \gamma_p) \cdots U (B, \beta_1) \, U (C, \gamma_1)\, |\psi(0)\rangle,
\label{eq:ansatz}
\end{equation}
where the evolution operators are given by $U (C,\gamma) = e^{-i\gamma C}$ and $U (B, \beta) = e^{-i\beta B}$. The integer $p$ is a parameter characterizing a variational ansatz.

For a given $p$, the goal of the algorithm is to find a set of
variational parameters that minimizes the expectation value of $ F_p
(\boldsymbol{\gamma}, \boldsymbol{\beta}) = \bra{\boldsymbol{\gamma},
  \boldsymbol{\beta}} C \ket{\boldsymbol{\gamma},
  \boldsymbol{\beta}}$, which ensures that the state
$\ket{\boldsymbol{\gamma}, \boldsymbol{\beta}}$ approaches the ground
state of $C$. Physically, the ansatz describes time evolution for a
total time $ T=\sum_{i=1}^p\left(\gamma_i+\beta_i\right)$, and a
sequence of sudden switching between the Hamiltonians $B$ and
$C$. While this ansatz with a finite $p$ is an intelligent guess,
the result that we derived using Pontryagin's minimum principle implies
that given bounded independent control of Hamiltonian terms, the ansatz (\ref{eq:ansatz}) is the optimal
choice for a VQA approach to optimization.  We reiterate that $B$ and $C$ are each a sum of commuting one- and/or two-qubit terms. Therefore, our protocol can be interpreted as a sequence of simple gates. Estimating the required $p$ requires an
analysis of the characteristic time scales of the pulses, which we
carry out in this paper.

\section{Numerical studies}
\label{sec:numerical}
We start by verifying for small system sizes and short annealing times that the optimal annealing protocol is indeed bang-bang, by using a Metropolis Monte Carlo (MC) algorithm, which makes no assumptions about the nature of the protocol. We divide the total time $T$ into $S$ slices of duration $\delta t=T/S$ and use a piece-wise constant protocol. The method approaches an unbiased optimization, i.e., it explores all permissible controls and chooses the optimal one, if the protocols obtained converge upon increasing $S$. We then proceed by carrying out a MC simulation starting from random initial protocols, without any assumption regarding the bang-bang nature of the protocol.  In each step, we slightly change the control parameter $g$ in a randomly chosen discretized time interval. If the cost function gets smaller, we accept the attempt; if the cost function gets larger, we accept the attempt with probability ${\rm e}^{-\Delta E/T_{MC}}$, where $T_{MC}$ is a fictitious temperature that is gradually reduced to zero. 

\begin{figure}[hbt]
\centering
\includegraphics[angle=0,origin=c,width=8.5cm]{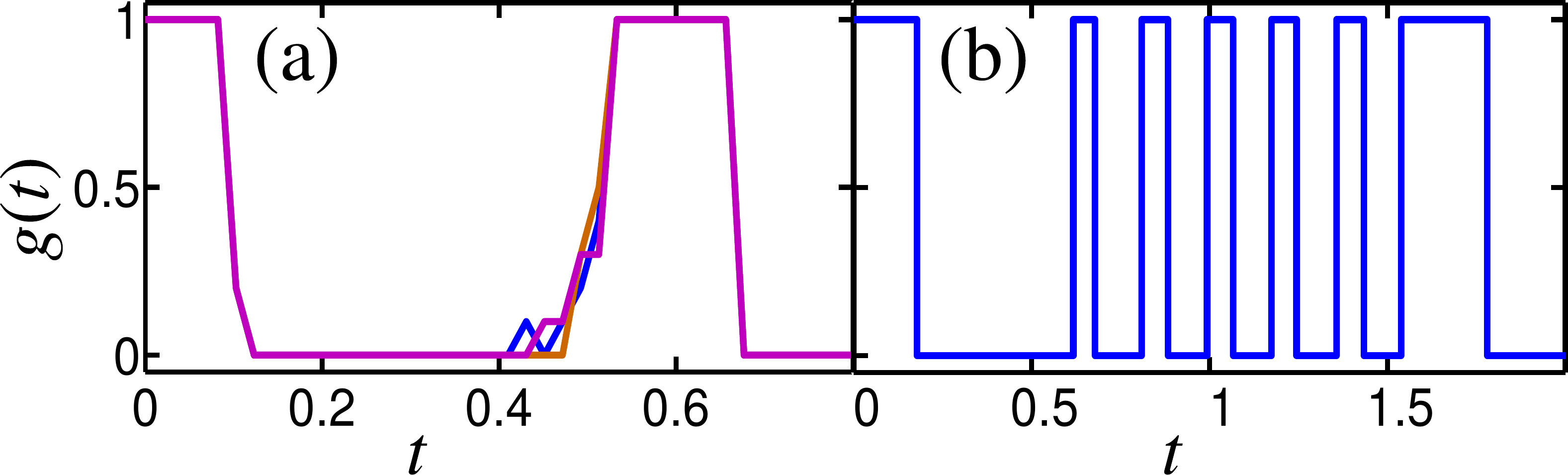}
\caption{(a) The optimal protocol obtained from MC simulations for a fixed instance of Hamiltonian (\ref{eq1}) with $n=5$ spins and total annealing time $T=0.8$. Different colors represent different initial protocols. The plots are for $S=40$, but the optimal protocol does not change upon increasing $S$. (b) A typical protocol obtained for a given instance of Hamiltonian (\ref{eq1}) with $n=5$ spins and $T=2$, using a classical optimization solver. We start from a uniform initial protocol with $S$ slices such that $\delta t=T/S=0.1$.}
\label{fig:protocol}
\end{figure}

In Fig. \ref{fig:protocol}(a), we show the optimal protocol obtained from such MC simulation for a fixed instance of Hamiltonian (\ref{eq1}) with $n=5$ spins and total time $T=0.8$.

Indeed, the MC simulation converge to a bang-bang protocol for different initial protocols in agreement with the Pontryagin's minimum principle.
Despite the convergence for short total time, the MC simulations often fail to converge for longer times and larger systems, signaling the difficulty of implementing VQA without any a priori knowledge about the form of the optimal protocol. However, based on the mathematical proof of the bang-bang nature of the optimal protocols, we can parameterize the protocol similar to QAOA \cite{Farhi} and use the durations of the pulses as variational parameters to be optimized with the interior-point minimization method (IPMM), increasing $p$ to achieve convergence.

We have checked that IPMM results are indeed  in agreement with MC results, e.g., Fig. \ref{fig:protocol}(a) (it also runs much faster). In Fig.\ref{fig:protocol}(b), we show a typical optimized protocol obtained with IPMM for a certain instance of Hamiltonian (\ref{eq1}) with $n=5$ spins and $T=2$. Guided by MC results, we choose around $\sim 20\times T$ variational parameters, which proved to be adequate (we converged to a smaller number of bangs than we allowed in the ansatz).

\begin{figure}[!hbt]
\centering
\includegraphics[angle=0,origin=c,width=8cm]{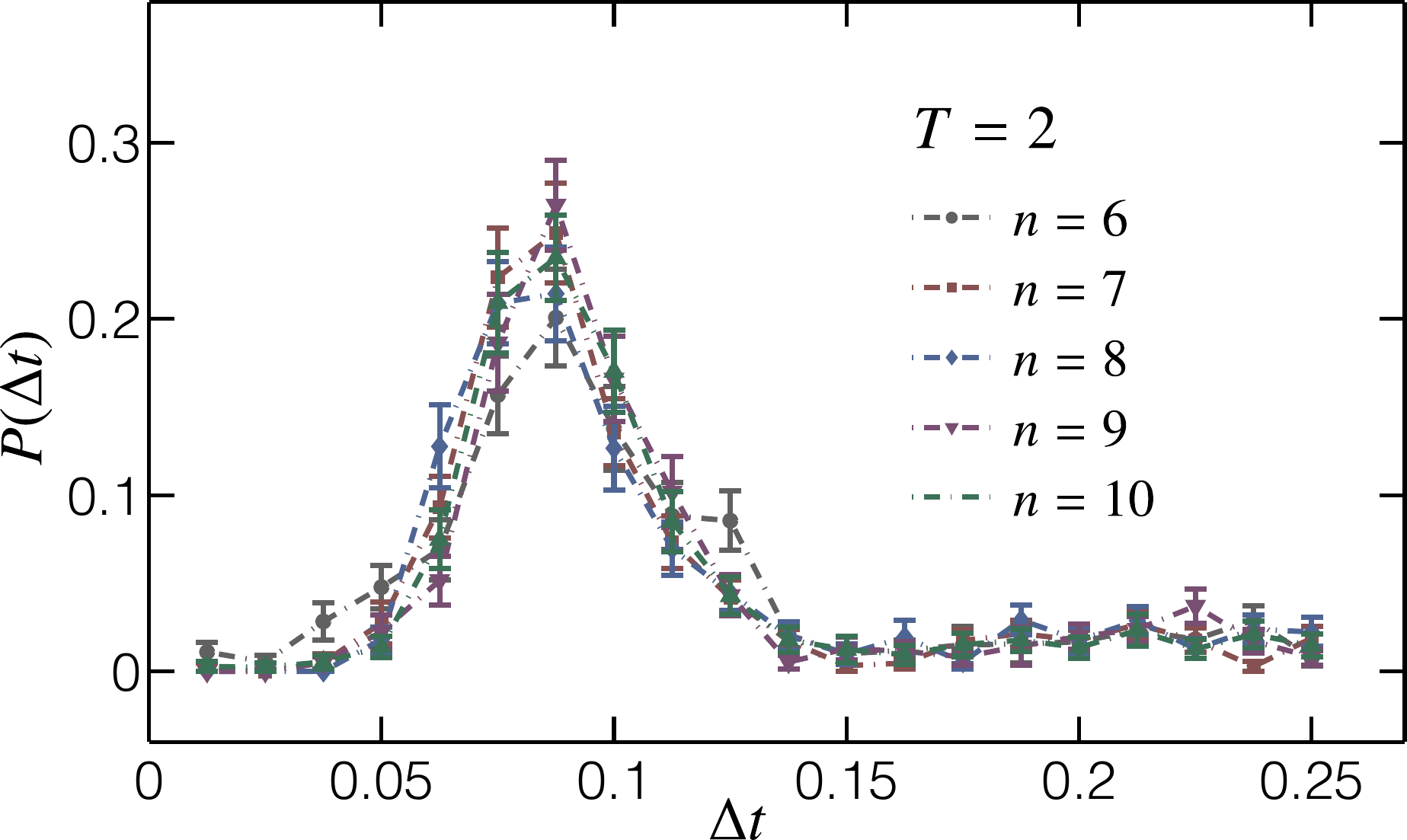}
\caption{The average probability distribution of the time scales of bangs for different system sizes: $n=6, 7, 8, 9$ and 10. The total annealing time is fixed to be $T=2$, leading to an average success rate around $0.33\sim0.47$, depending on the system sizes. Each curve is averaged over 50 instances of Hamiltonian (\ref{eq1}).}
\label{fig:distribution}
\end{figure}

We now turn to the critical question of the time scales of the pulses. We observe numerically and then argue analytically that the typical time scale of each bang is independent of the system size, and is only determined by some characteristic energy scale of Hamiltonian (\ref{eq1}). Therefore, from a complexity theory perspective, this result implies that the hardness of the optimization problem should translate into the number of pulses and/or the hardness of the protocol optimization. In Fig. \ref{fig:distribution} we plot the distribution of the time scales of each bang $\Delta t$ for system sizes $n=6, 7, 8, 9$ and 10. For each system size, we fix the total annealing time to be $T=2$, and average over 50 instances of the Hamiltonian (\ref{eq1}). 

We find that the distributions for the bang time collapse for different system sizes, and peak at almost the same value. This observation suggests a universal average distribution of the bang times for the near optimal protocols, and a typical time scale (peak or average value) that is independent of the system size. Although we have only considered a few system sizes, the dependence on $n$ is extremely weak and we expect our results to extrapolate to large $n$.

Finally we comment on the performance of our protocols. The cost function we minimize is the expectation value $\langle \psi(T) |C|\psi(T)\rangle$. Minimizing the energy expectation value results in larger overlap with the ground state of $C$. As expected, the time scales for our protocols are significantly shorter than those of the adiabatic algorithm with similar success rate. A comparison between the optimal bang-bang protocols and linear ramps $g(t)=t/T$ is shown in Fig. \ref{fig:noise}. {The errors in the final wavefunctions $1-|\langle\psi_{GS}|\psi(T)\rangle|^2$ and final energies $E-E_{GS}\equiv\langle \psi(T) |C|\psi(T)\rangle-E_{GS}$ are averaged over the 20 instances (out of 50 generated realizations) with the highest success rates, for the optimal bang-bang protocol and linear QAA ramp respectively. We find that, in the system sizes considered in this work, the nonadiabatic bang-bang protocol with the same total time, significantly outperforms the linear ramp (commonly used in QAA) in the ideal case, where the thermal environment and external noise are neglected. Of course, in practice one needs to include the overhead of searching for the optimal solution, and understand how it scales as function of system size. In particular, while implementing the bang-bang protocol consisting of square pulses on a quantum annealer is feasible~\cite{Barends}, finding the optimal protocol with a classical optimizer could be difficult for certain hard instances of problems.

\begin{figure}[b]
\centering
\includegraphics[angle=0,origin=c,width=9cm]{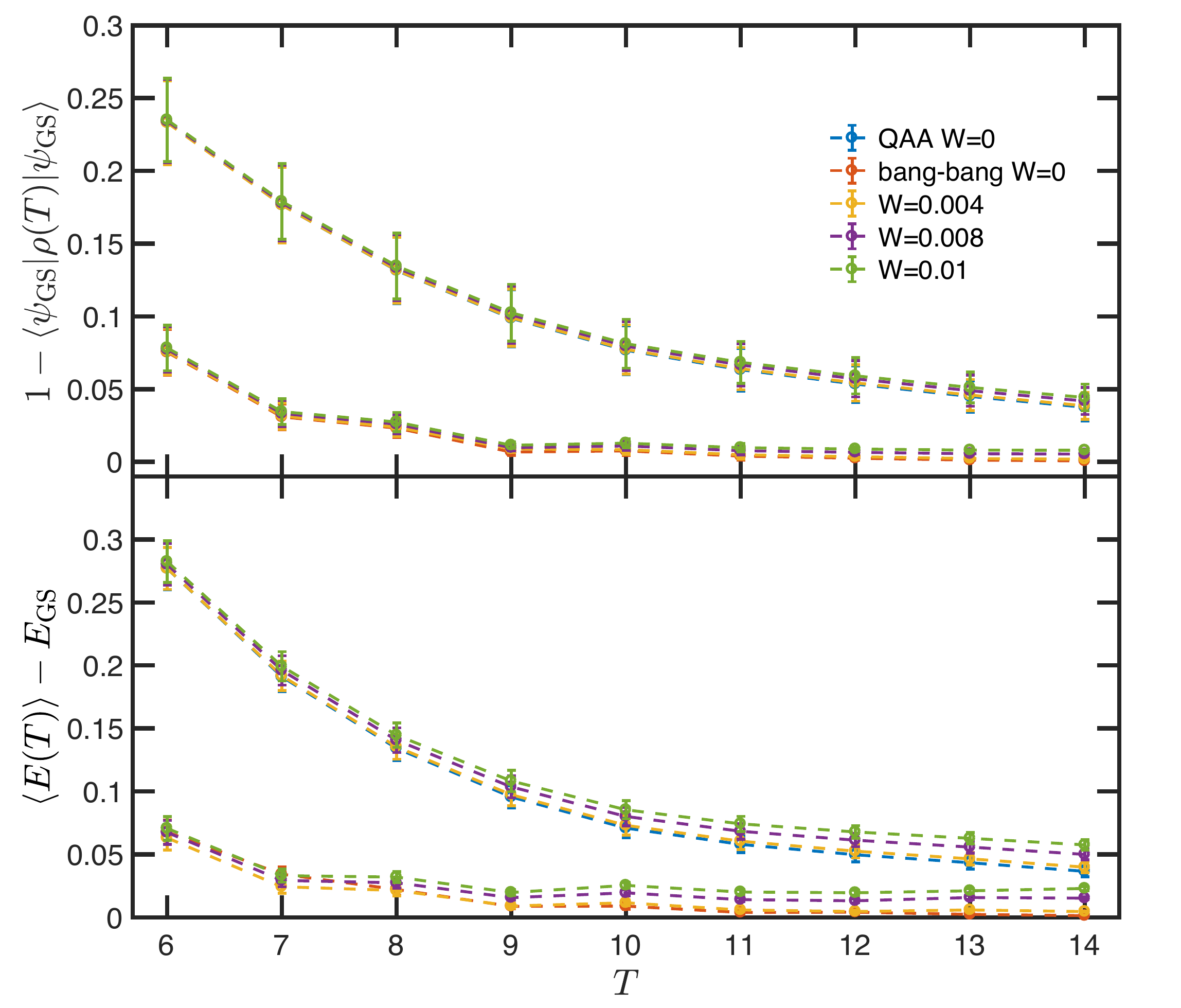}
\caption{Errors in the fidelity (upper panel) and final energies (lower panel) evolved with the bang-bang and QAA protocols in the presence of noise with different strengths for $n=5$.}
\label{fig:noise}
\end{figure}

\section{Effects of Dissipation and Dephasing}
\label{sec:noise}

Real-world implementations of the bang-bang and QAA protocols are inevitably subject to noise either in the external controls or due to coupling to the thermal environment. Therefore, it is important to examine the effects of these perturbations on our optimal protocols for practical applications. Here we consider two noise models in order to evaluate the robustness of our bang-bang protocol, at the same time comparing it with the performance of QAA.

\subsection{Random Dephasing Noise}

Here we consider pure dephasing noise, where we introduce random fields in the $x$ and $z$ directions. This type of noise can capture noise induced by hardware electronics. Our error model can be viewed as the continuous time analog of the depolarizing channel commonly used for simulating noise in quantum circuits \cite{Nielson}. Since in a bang-bang protocol we either have $g(t)=1$ or $g(t)=0$ at any given time, we can write the stochastic Hamiltonian as
\begin{eqnarray}
H(t)&=& C+\sum_{i=1}^n \delta h_i (t)\sigma^z_i+\sum_{i=1}^n \delta b_i(t)\sigma^x_i,\quad g=1,\\
H(t) &=& B+\sum_{i=1}^n \delta h_i (t)\sigma^z_i+\sum_{i=1}^n \delta b_i(t)\sigma^x_i,\quad g=0,
\end{eqnarray}
where $\delta h_i(t)$ and $\delta b_i(t)$ are noise in the $z$ and $x$ directions respectively, with strengths independent of the value of the coupling constants (additive noise). Assuming independent white noise for different terms with zero mean and second moments
\begin{eqnarray}
\label{whitenoise1}
\overline{\delta h_{i}(t)\delta h_{i'}(t')}&=&W_h^2\delta_{ii'}\delta(t-t'),\\
\overline{\delta b_{i}(t)\delta b_{i'}(t')}&=&W_b^2\delta_{ii'}\delta(t-t'),
\label{whitenoise2}
\end{eqnarray}
the noise-averaged density matrix evolves with the following master equation \cite{Rahmani2015}
\begin{equation}
\begin{split}
{d\rho(t)\over dt}= &-i[H,\rho(t)]-{1\over 2}W_h^2\sum_{i=1}^n[[\rho(t),\sigma^z_i],\sigma^z_i]\\
&-{1\over 2}W_b^2\sum_{i=1}^n[[\rho(t),\sigma^x_i],\sigma^x_i],
\label{eqn:master2}
\end{split}
\end{equation}
where we take $W_b=W_h=W$ for simplicity. In the bang-bang case, the Hamiltonian $H$ takes two different values $H=C$ ($H=B$) for $g=1$ ($g=0$), while in the QAA case, $H$ has the explicit time dependence of Eq.~\eqref{eq2} with $g(t)=t/T$.

In Fig.~\ref{fig:noise} we show the errors in the fidelity $1-\langle
\psi_{GS}|\rho(T)|\psi_{GS}\rangle$ and final energy ${\rm
  Tr}\left[\rho(T) C\right]-E_{GS}$ for different strengths of
noise. We find that in the small $W$ regime, the noise only slightly
decreases the fidelity, acting like a perturbation without inducing
any instability. The effects of the noise on the linear QAA ramp are
similar both qualitatively and quantitatively, changing the $W=0$
error by amount of the same order of magnitude. For the strongest
strength of noise that we studied ($W=0.01$), the fidelity of the
optimal bang-bang protocol remains higher than that of the linear ramp
protocol.

A comment is in order regarding the dimension of $W$ and the range used. As $\delta(t-t')$ has a dimension of time (inverse energy), $W^2$ has a dimension of energy. Strictly speaking, the $\delta$ function introduces infinitely large (albeit completely uncorrelated) random fields. This is unrealistic. In real experiments there is a characteristic high frequency, introducing a characteristic short time scale $\Delta \tau$, over which noise is correlated. This frequency scale is typically several orders of magnitude larger than the characteristic energy of the Hamiltonian (it diverges for the $\delta$ function). Therefore, Eq. (\ref{whitenoise1}) and (\ref{whitenoise2}) imply that $\delta h, \delta b \sim W/\sqrt{\Delta \tau}$, which means that for moderate noise in the random fields $\delta h$ and $\delta b$, the corresponding values of $W$ are suppressed by $\sqrt{\Delta \tau}$.

\subsection{Weak Thermal Bath}

Here we consider coupling the system to a weak thermal bath at temperature $1/\beta$. In this regime, the dynamics of the open system can be approximately described by the Redfield master equation which is commonly used to model noisy QAA for an actual annealing hardware \cite{Amin, Albash, Boixo}. Here we use the formulation in Ref. \cite{Amin} and apply it to both QAA and bang-bang protocols.

The system of many qubits is coupled to the thermal bath via the
Hamiltonian $\sum_i^{n} \sigma^z_i Q^z_i$, where $Q^z_i$ are bath
operators.  We assume an Ohmic bosonic bath in thermal equilibrium,
with the spectral density function given by
\begin{eqnarray}
  S^z_i(\omega)=\int_{-\infty}^{\infty} dt \;e^{i\omega t} \;\langle Q^z_i(t) Q^z_i(0)\rangle = \eta \omega\; \frac{1}{1-e^{-\beta\omega}}
  \;,
 \end{eqnarray}
where $\eta$ is a dimensionless coefficient describing the strength of
the coupling to the environment. We have taken the cut-off frequency
of the bath to be infinite, so as to guarantee the Markovian
assumption of dynamics. We employed Eqs.(4-9) in reference \cite{Amin}
to simulate the dynamics of open systems based on the Redfield master
equation.

\begin{figure}[h]
\centering
\includegraphics[angle=0,origin=c,width=9cm]{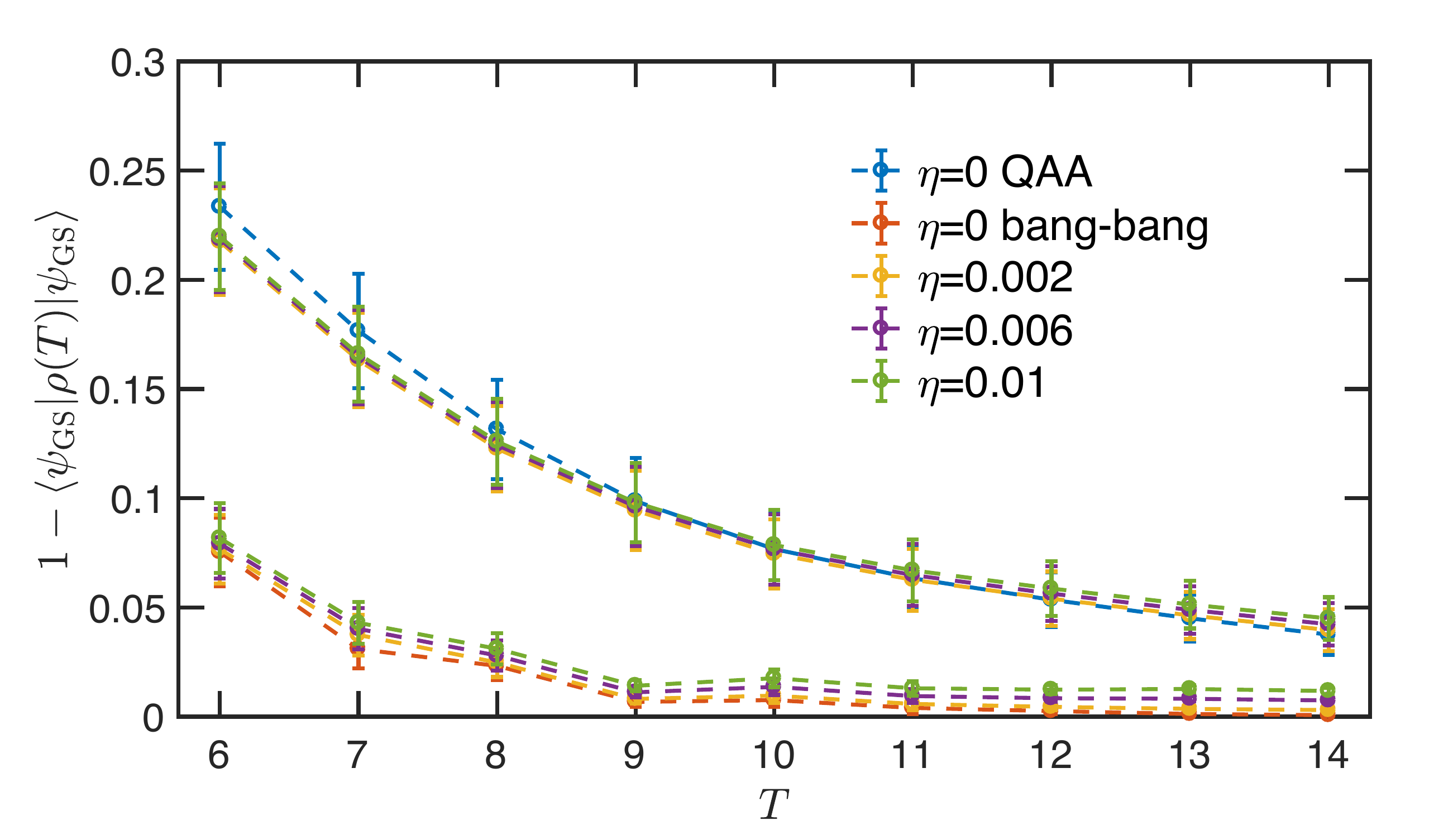}
\caption{Errors in the fidelity of final states evolved with the bang-bang and QAA protocols in the presence of different strengths of coupling to the environment for $n=5$. The inverse temperature is chosen to be $\beta=2/J$.}
\label{fig:redfield}
\end{figure}

Fig.~\ref{fig:redfield} shows the errors in the fidelity for different
strengths of coupling to the bath, for both QAA and bang-bang
protocols. Similar to the case of the closed system in the presence of
white noise, we find that the errors corresponding to both protocols
change in an analogous manner due to weak coupling to the environment in
both the short- and long-time regimes. There is an intermediate time
regime $8.5\lesssim T\lesssim11.5$, where QAA exhibits remarkable
robustness and a much smaller change in $\eta=0$ error. However, the
errors of the VQA and QAA get closer as we increase $T$. Once again,
the fidelity of the optimal bang-bang protocol remains higher than the
QAA even for open system dynamics.

\section{Pulse duration from the Pontryagin's minimum principle}

Here we provide more details on how the Pontryagin's minimum principle can not only tell about the form of optimal solution for VQA but can also shed light on when the pulses should be switched on and off, in the context of the SK model.

Using the computational basis $z=z_1\dots z_n$, we represent the wave function as $\ket{\psi(t)}=\sum_z A_z(t)\ket{z}$.
The initial state with all the spins in the $x$ direction corresponds to $A_z(0)= 1/{\sqrt{2^n}}$, and the Schr\"odinger equation reads 
 \begin{equation}
i\partial_t A_z(t)=g(t)C_zA_z(t)+[1-g(t)]\sum_{k=1}^n A_{\bar {z}(k)}(t),
\end{equation}
with $\bar {z}(k)=z_1\dots \bar{z}_k\dots z_n$, where $\bar{z}_k$ represents a flipped spin with respect to $z_k$ and $C_z$ is the energy function we would like to minimize.  In terms of the real and imaginary parts of $A_z(t)=R_z(t)+i I_z(t)$ , we can then write
\begin{eqnarray}
\partial_t R_z(t)&=&g(t)C_zI_z(t)+[1-g(t)]\sum_{k=1}^n I_{\bar {z}(k)}(t),\\
\partial_t I_z(t)&=&-g(t)C_zR_z(t)-[1-g(t)]\sum_{k=1}^n R_{\bar {z}(k)}(t).
\end{eqnarray}
Introducing conjugate momenta $P_z(t)$ and $Q_z(t)$ respectively for the real and imaginary parts of $A_z(t)$, the explicit form of the optimal-control Hamiltonian is given by
 \begin{equation}
 \begin{split}
{\cal H}=\sum_{z}\bigg\{&g(t)C_z\left[P_z(t)I_z(t)-Q_z(t)R_z(t)\right]\\
&+[1-g(t)]\sum_{k=1}^n \left[P_z(t)I_{\bar {z}(k)}(t)-Q_z(t)R_{\bar {z}(k)}(t)\right]\bigg\}.
\end{split}
\end{equation}

 The equations of motion for the conjugate momenta are $\partial_t P_z(t)=-{\partial{\cal H}\over \partial R_z(t)}$ and $\partial_t Q_z(t)=-{\partial{\cal H}\over \partial I_z(t)}$, which can be written explicitly as
 \begin{eqnarray}
\partial_t P_z(t)&=&g(t)C_zQ_z(t)+[1-g(t)]\sum_{k=1}^n Q_{\bar {z}(k)}(t),\\
\partial_t Q_z(t)&=&-g(t)C_zP_z(t)-[1-g(t)]\sum_{k=1}^n P_{\bar {z}(k)}(t),
\end{eqnarray}
where we have used the relationships $\sum_{z,k}Q_z(t)R_{\bar {z}(k)}(t)=\sum_{z,k}Q_{\bar {z}(k)}(t)R_z(t)$ and $\sum_{z,k}P_z(t)I_{\bar {z}(k)}(t)=\sum_{z,k}P_{\bar {z}(k)}(t)I_z(t)$.

The cost function
 \begin{equation}
F[g(t)]=\sum_z|A_z(T)|^2C_z
\end{equation}
leads to the following boundary conditions at $t=T$ for the conjugate momenta:
 \begin{equation}
P_z(T)=2R_z(T)C_z,\quad Q_z(T)=2I_z(T)C_z.
\label{eqn:bc}
\end{equation}
Notice that $g(t)$ uniquely determines $A_z(t)$. From $A_z(T)$ and the expression above, we can find $P_z(T)$ and $Q_z(T)$, solve the equations of motion backward in time and determine the conjugate momenta as a function of time. Therefore, $g(t)$ also uniquely determines $P_z(t)$ and $Q_z(t)$. The Potryagin's minimum principle states that the optimal protocol $g^*(t)$ satisfies
 \begin{equation}
{\cal H}(g^*,{\bf R}^*,{\bf I}^*,{\bf P}^*,{\bf Q}^*)=\min_{g}{\cal H}(g,{\bf R}^*,{\bf I}^*,{\bf P}^*,{\bf Q}^*),
\end{equation}
where ${\bf R}^*,{\bf I}^*,{\bf P}^*,{\bf Q}^*$ are the corresponding optimal solution. As argued in Sec.~\ref{sec:pontryagin}, $g^*(t)$ is bang-bang and can only take two values of $0$ and $1$. Which value will depend on the sign of $\partial{\mathcal H}\over \partial g$ given by the expression
 \begin{equation}
 \begin{split}
{\partial{\mathcal H}\over \partial g}=\sum_{z}\bigg\{&C_z\left[P^*_z(t)I^*_z(t)-Q^*_z(t)R^*_z(t)\right]\\
&-\sum_{k=1}^n \left[P^*_z(t)I^*_{\bar {z}(k)}(t)-Q^*_z(t)R^*_{\bar {z}(k)}(t)\right]\bigg\}.
\end{split}
\label{eq:coeff}
\end{equation}

The sudden quenches at which $g(t)$ switches from 0 to 1 or vice versa correspond to the zeros of $\partial{\mathcal H}\over \partial g$ above. 

Let us first combine $P_z$ and $Q_z$ into one complex momentum
\begin{equation}
\Pi_z(t)=P_z(t)+iQ_z(t).
\end{equation}
As we argued above, the optimal protocol is bang-bang with  $g(t)=0,1$. In any interval with $g(t)=1$, we can write
\begin{eqnarray}
A_z(t)&=& {\rm e}^{-iC_z(t-t_0)}A(t_0) \\
\Pi_z(t)&=& {\rm e}^{-iC_z(t-t_0)}\Pi(t_0),
\end{eqnarray}
where $t_0$ is the beginning of the current bang $g(t_0)=1$. We first note that the term $P_z(t)I_z(t)-Q_z(t)R_z(t)={\rm Im}\left[A_z(t)\Pi^*_z(t)\right]$ in Eq.~\eqref{eq:coeff} and does not change in intervals with $g(t)=1$. Moreover the terms
\begin{equation}
\begin{split}
P_z(t)&I_{\bar {z}(k)}(t)-Q_z(t)R_{\bar {z}(k)}(t)\\
=&{\rm Im}\left[{\rm e}^{-i\left(C_z-C_{\bar {z}(k)}\right)(t-t_0)}A_z(t_0)\Pi^*_{\bar {z}(k)}(t_0)\right].
\end{split}
\end{equation}

 The above equation allows for an estimation of the typical time scale of the interval $g(t)=1$. Suppose at some $t_0$, $g(t)$ switches from 0 to 1, i.e. a bang starts. From the discussion above, we know that at $t=t_0$ we must have ${\partial \mathcal{H}\over \partial g}(t_0)=0$. The time when the bang stops corresponds to the next $t>t_0$ when ${\partial \mathcal{H} \over \partial g}(t)=0$. More explicitly, in the interval $g(t)=1$, Eq. (\ref{eq:coeff}) can be written as
 \begin{equation}\label{eqn20}
w(t)=\sum_z\sum_{k=1}^n {\rm Im}\left [ ({\rm e}^{-i(C_z-C_{\bar{z}(k)})(t-t_0)}-1)A_z(t_0)\Pi_{\bar{z}(k)}^*(t_0)\right].
\end{equation}
We have $w(t_0)=0$. The first root of the equation $w(t)=0$ with $t>t_0$ then determines the duration of a pulse.

While we cannot derive analytically the average first root for $t>t_0$ from Eq.~\eqref{eqn20}, one can see that the only time dependence in Eq. (\ref{eqn20}) within the current interval is in ${\rm
  e}^{-i(C_z-C_{\bar{z}(k)})t}$. Thus, the energy difference $\Delta
C_{z,k}\equiv C_z-C_{\bar{z}(k)}$, which has zero mean (as both $C_z$ and $C_{\bar{z}(k)}$ have zero mean) and variance
$\overline{\Delta C_{z,k}^2}=4(J^2+h^2)$, sets the characteristic time
scale proportional to $1/\sqrt{J^2+h^2}$ observed in
Fig.~\ref{fig:distribution}. Importantly, this time scale is finite and system-size independent,
distinguishing our bang-bang type optimal protocol from the
Trotterization of generic protocols, where the duration of
individual pulses must be taken to zero. We believe the numerically obtained system-size independent distribution of Fig.~\ref{fig:distribution} follows from Eq.~\eqref{eqn20} whose root determines one set of the switching times for quenching $g(t)$ from 1 to 0, giving the duration of a pulse with $g(t)=1$ [the distributions of pulse durations with $g(t)=0$ and $g(t)=1$ were found numerically to be almost identical]. However, an analytical derivation of the distribution in Fig.~\ref{fig:distribution} has remained elusive.

\section{Summary and Outlook} We have shown that the optimal VQA
with bounded linear control parameters has a protocol of the bang-bang form. We
verified this prediction by finding numerically the optimal protocol
that minimizes the energy of a SK spin glass. The optimal nonadiabatic bang-bang protocols significantly reduce the error when compared to QAA within the same running time, and, at least for our system sizes, the advantage remains in the presence of weak additive white noise in the control parameters as well as weak coupling to a thermal environment.

Importantly, we
show that the characteristic time scale between bangs is fixed by
the energy scales in the problem and is independent of system size,
which we confirm numerically. This finding significantly reduces the number of variational parameters in VQA, potentially decreasing the computational cost of the VQA outer-loop classical optimization algorithm to a great extent. 

Our results, that the bang-bang protocols are optimal and the duration of each square pulse is size-independent, inform the search for effective hybrid classical-quantum schemes for solving combinatorial optimization problems. Further progress relies on the development of efficient outer-loop algorithms as well as hardware development for quantum enhanced optimization. Ultimately, the power of our results lies in their application to larger systems, for which solving the time-dependent Schrodinger equation is impossible on a classical computer. Rapid developments in quantum technologies~\cite{Mohseni17}, together with the relative robustness of our protocols to specific models of external noise and thermal environment support the promise of such applications.

\acknowledgements{We are grateful to Ryan Babbush, Pedram Roushan, Eduardo Mucciolo, Dries Sels and Anatoli Polkovnikov for illuminating discussions. Z.-C. Y. would like to thank Quntao
  Zhuang for many early discussions on QAA. This work was supported by
  NSERC (A. R.), Max Planck-UBC Centre for Quantum Materials (A. R.), and DOE
  Grant No. DE-FG02-06ER46316 (C. C. and Z.-C. Y.).}


\begin{thebibliography}{99}

\bibitem{Nishimori} T. Kadowaki and H. Nishimori, {\it Quantum annealing in the transverse Ising model}, 
Phys. Rev. E \textbf{58}, 5355 (1998).

\bibitem{Sipser} E. Farhi, J. Goldstone, S. Gutmann, and M. Sipser, {\it Quantum Computation by Adiabatic Evolution}, arXiv:quant-ph/0001106.

\bibitem{QAA} E. Farhi, J. Goldstone, S. Gutmann, J. Laplan, A. Lundgren, and D. Preda, {\it A Quantum Adiabatic Evolution Algorithm Applied to Random Instances of an NP-Complete Problem}, Science {\bf 292}, 472 (2001).

\bibitem{Perdomo} A. Perdomo-Ortiz, S. E. Venegas-Andraca, and A. Aspuru-Guzik, {\it A study of heuristic guesses for adiabatic quantum computation}, Quantum Info. Process. {\bf 10}, 33 (2011).

\bibitem{Mohan1} L. Zeng, J. Zhang, and M. Sarovar, {\it Schedule path optimization for adiabatic quantum computing and optimization}, J. Phys. A: Math. Theor. {\bf 49}, 165304 (2016).

\bibitem{zhuang} Q. Zhuang, {\it Increase of degeneracy improves the performance of the quantum adiabatic algorithm}, Phys. Rev. A. {\bf 90} (5), 052317 (2014). 

\bibitem{Rezakhani} A. T. Rezakhani, W.-J. Kuo, A. Hamma, D. A. Lidar, and P. Zanardi, {\it Quantum Adiabatic Brachistochrone}, Phys. Rev. Lett. {\bf 103}, 080502 (2009).

\bibitem{troyer} D. S. Steiger, T. F. R{\o}nnow, and M. Troyer, {\it Heavy Tails in the Distribution of Time to Solution for Classical and Quantum Annealing}, Phys. Rev. Lett. {\bf 115}, 230501 (2015).

\bibitem{heim} B. Heim, T. F. R{\o}nnow, S. V. Isakov, and M. Troyer, {\it Quantum versus classical annealing of Ising spin glasses}, Science {\bf 348}, 215 (2015).

\bibitem{crosson} E. Crosson, E. Farhi, C. Y. Lin, H. H. Lin, and P. Shor, {\it Different Strategies for Optimization Using the Quantum Adiabatic Algorithm}, arXiv:1401.7320.

\bibitem{Rohringer}
W. Rohringer, R. B\"ucker, S. Manz, T. Betz, C. Koller, M. Gbel, A. Perrin, J. Schmiedmayer, and T. Schumm, {\it Stochastic optimization of a cold atom experiment using a genetic algorithm}, Appl. Phys. Lett. \textbf{93}, 264101 (2008).

\bibitem{Rosi}
S. Rosi, A. Bernard, N. Fabbri, L. Fallani, C. Fort, M. Inguscio, T. Calarco, and S. Montangero,  {\it Fast closed-loop optimal control of ultracold atoms in an optical lattice}, Phys. Rev. A \textbf{88}, 021601 (2013).

\bibitem{Rahmani1}
A. Rahmani, T. Kitagawa, E. Demler, and C. Chamon, {\it Cooling through optimal control of quantum evolution}, Phys. Rev. A \textbf{87}, 043607 (2013).


\bibitem{Rahmani2}
A. Rahmani, {\it Quantum Dynamics with AN Ensemble of Hamiltonians}, Mod. Phys. Lett. B \textbf{27}, 1330019 (2013).


\bibitem{Peruzzo:14} A. Peruzzo, J. McClean, P. Shadbolt, M. H. Yung, X. Q. Zhou, P. J. Love, A. Aspuru-Guzik, and J. L. O'Brien, {\it A variational eigenvalue solver on a photonic quantum processor}, Nat. Comm. {\bf5}, 4213, (2014).

\bibitem{Yung:14} M. Yung, J. Casanova, A. Mezzacapo, J. McClean, L. Lamata, A. Aspuru-Guzik, and E. Solano, {\it From transistor to trapped-ion computers for quantum chemistry}, Scientific Reports, {\bf4}:3589 (2014)

\bibitem{Wecker:15} D. Wecker, M. B. Hastings, and M. Troyer, {\it Progress towards practical quantum variational algorithms}, Phys. Rev. A {\bf92}, 042303 (2015).

\bibitem{wecker} D. Wecker, M. B. Hastings, and M. Troyer, {\it Training a quantum optimizer}, Phys. Rev. A {\bf94}, 022309 (2016).

\bibitem{McClean:16} J. R. McClean, J. Romero, R. Babbush, and A. Aspuru-Guzik, {\it The theory of variational hybrid quantum-classical algorithms}, New J. Phys. {\bf18}, 023023 (2016).


\bibitem{McClean:16-2} Jarrod R. McClean, Mollie E. Schwartz, Jonathan Carter, and Wibe A. de Jong, {\it Hybrid Quantum-Classical Hierarchy for Mitigation of Decoherence and Determination of Excited States},  arXiv:1603.05681.


\bibitem{Shen:15} Y. Shen, X. Zhang, S. Zhang, J. N. Zhang, M. H. Yung, and K. Kim, {\it Quantum implementation of the unitary coupled cluster for simulating molecular electronic structure}, Phys. Rev. A {\bf95}, 020501(R) (2017).

\bibitem{Eichler:15} C. Eichler, J. Mlynek, J. Butscher, P. Kurpiers, K. Hammerer, T. J. Osborne, and A. Wallraff, {\it Exploring Interacting Quantum Many-Body Systems by Experimentally Creating Continuous Matrix Product States in Superconducting Circuits}, Phys. Rev. X {\bf5}, 041044 (2015).

\bibitem{Omalley:15} P. O'Malley, R. Babbush, I. Kivlichan, J. Romero, J. McClean, R. Barends, J. Kelly, P. Roushan, A. Tranter, N. Ding, B. Campbell, Y. Chen, Z. Chen, B. Chiaro, A. Dunsworth, A. Fowler, E. Jeffrey, A. Megrant, J. Mutus, C. Neill, C. Quintana, D. Sank, A. Vainsencher, J. Wenner, T. White, P. Coveney, P. Love, H. Neven, A. Aspuru-Guzik, and J. Martinis., {\it Scalable Quantum Simulation of Molecular Energies}, Phys. Rev. X \textbf{6}, 031007 (2016).

\bibitem{Farhi} E. Farhi, J. Goldstone, and S. Gutmann, {\it A Quantum Approximate Optimization Algorithm}, arXiv:1411.4028.

\bibitem{Farhi:2} E. Farhi, J. Goldstone, and S. Gutmann, {\it A Quantum Approximate Optimization Algorithm Applied to a Bounded Occurrence Constraint Problem}, arXiv:1412.6062.

\bibitem{Farhi:3} E. Farhi, and A. W. Harrow, {\it Quantum Supremacy through the Quantum Approximate Optimization Algorithm}, arXiv:1602.07674.

\bibitem{Yechao} C. Y.-Y Lin, and Y. Zhu, {\it Performance of QAOA on Typical Instances of Constraint Satisfaction Problems with Bounded Degree}, arXiv:1601.01744.

\bibitem{control1} A. E. Bryson, and Y.-C. Ho, \textit{Applied optimal control: optimization, estimation and control} (CRC Press, 1975).

\bibitem{Pontryagin} L. S. Pontryagin, \textit{Mathematical Theory of Optimal Processes} (CRC Press, 1987).

\bibitem{control2} R. F. Stengel, \textit{Optimal control and estimation} (Courier Corporation 2012).

\bibitem{Mohan2} C. Brif, M. D. Grace, M. Sarovar, and K. C. Young, {\it Exploring adiabatic quantum trajectories via optimal control}, New J. Phys. {\bf 16}, 065013 (2014).

\bibitem{adaptive} K. J. Astrom, {\it Adaptive feedback control}, Proceedings of the IEEE  {\bf 75}(2), 185-217 (1987).

\bibitem{Brif:10} C. Brif, R. Chakrabarti, and H. Rabitz, {\it Control of quantum phenomena: past, present and future}, New J. Phys. {\bf12}, 075008 (2010).

\bibitem{Dong:10} D. Dong, and I. R. Petersen, {\it Quantum control theory and applications: a survey}, IET Control Theory A  {\bf12}, 2651, (2010).


\bibitem{Lucas:14} A. Lucas, {\it Ising formulations of many NP problems}, Frontiers in Physics {\bf2}, 5 (2014).

\bibitem{Rahmani3}
T. Karzig, A. Rahmani, F. von Oppen, and G. Refael, {\it Optimal control of Majorana zero modes}, Phys. Rev. B \textbf{91}, 201404 (2015).
 
\bibitem{Rahmani4}
A. Rahmani, B. Seradjeh, and M. Franz,  {\it Topologically Protected Non-Abelian Braiding through Optimal Diabatic Dynamics}, arXiv:1605.03611.

\bibitem{Pilcher2013} H. Pichler, J. Schachenmayer, A. J. Daley, and P. Zoller, {\it Heating dynamics of bosonic atoms in a noisy optical lattice}, Phys. Rev. A {\bf87}, 033606 (2013).
	
	
\bibitem{Rahmani2015} A. Rahmani, {\it Dynamics of noisy quantum systems in the Heisenberg picture: Application to the stability of fractional charge}, Phys. Rev. A \textbf{92}, 042110 (2015).

\bibitem{open} H. P. Breuer, and F. Petruccione, 
\newblock{{\it The Theory of Open Quantum Systems} (Oxford University Press, New York, 2002).}

\bibitem{SK} D. Sherrington, and S. Kirkpatrick, {\it Solvable Model of a Spin-Glass}, Phys. Rev. Lett. {\bf 35}, 1792 (1975).

\bibitem{Dutta}
A. Dutta, A. Rahmani, and A. del Campo,  {\it Anti-Kibble-Zurek Behavior in Crossing the Quantum Critical Point of a Thermally Isolated System Driven by a Noisy Control Field}, Phys. Rev. Lett. {\bf 117}, 080402 (2016).

\bibitem{Barends} R. Barends, A. Shabani, L. Lamata, J. Kelly, A. Mezzacapo, U. Las Heras, R. Babbush, A. G. Fowler, B. Campbell, Yu Chen, Z. Chen, B. Chiaro, A. Dunsworth, E. Jeffrey, E. Lucero, A. Megrant, J. Y. Mutus, M. Neeley, C. Neill, P. J. J. O'Malley, C. Quintana, P. Roushan, D. Sank, A. Vainsencher, J. Wenner, T. C. White, E. Solano, H. Neven, and John M. Martinis, {\it Digitized adiabatic quantum computing with a superconducting circuit}, Nature {\bf 534}, 222-226 (2016).

\bibitem{Nielson} M. A. Nielson, and I. L. Chuang, \textit{Quantum computation and quantum information} (Cambridge University Press 2010).

\bibitem{Amin} M. H. S. Amin, C. J. S. Truncik, D. V. Averin, {\it Role of single-qubit decoherence time in adiabatic quantum computation}, Phys. Rev. A {\bf 80}, 022303 (2009).

\bibitem{Albash} T. Albash, S. Boixo, D. A. Lidar, P. Zanardi, {\it Quantum adiabatic Markovian master equations}, New J. Phys {\bf 14}, 123016 (2012).

\bibitem{Boixo} S. Boixo, V. N. Smelyanskiy, A. Shabani, S. V. Isakov, M. Dykman, V. S. Denchev, M. H. Amin, A. Y. Smirnov, M. Mohseni, and H. Neven, {\it Computational multiqubit tunnelling in programmable quantum annealers}, Nat Commun. {\bf 7}, 10327 (2016).

\bibitem{Mohseni17} M. Mohseni, P. Read, H. Neven, S. Boixo, V. Denchev, R. Babbush, A. Fowler, V. Smelyanskiy, and J. Martinis, {\it Commercialize quantum technologies in five years}, Nature \textbf{543}, 171 (2017).
\end{thebibliography}
\end{document}